\documentclass[letterpaper,aps,preprint,superscriptaddress]{revtex4}
\usepackage{graphicx}% Include figure files
\usepackage{dcolumn}% Align table columns on decimal point
\usepackage{bm}% bold math
\usepackage{amsmath}
\usepackage{xspace}
\usepackage{multirow}
\usepackage{natbib}
\usepackage{color}

\begin{document}

\preprint{}

\title{Observation of a novel orbital selective Mott transition in Ca$_{1.8}$Sr$_{0.2}$RuO$_4$}

\author{M. Neupane}
\affiliation{Department of Physics, Boston College, Chestnut Hill, MA 02467, USA}
\author{P. Richard}
\affiliation{Department of Physics, Boston College, Chestnut Hill, MA 02467, USA}
\affiliation{WPI-AIMR, Tohoku University, Sendai 980-8578, Japan}
\author{Z.-H. Pan}
\affiliation{Department of Physics, Boston College, Chestnut Hill, MA 02467, USA}
\author{Y. Xu}
\affiliation{Department of Physics, Boston College, Chestnut Hill, MA 02467, USA}
\author{R. Jin}
\affiliation{Condensed Matter Science Division,Oak Ridge National Laboratory,Oak Ridge, Tennessee 37831,USA}
\author{D. Mandrus}
\affiliation{Condensed Matter Science Division,Oak Ridge National Laboratory,Oak Ridge, Tennessee 37831,USA}
\author{X. Dai}
\affiliation{Beijing National Laboratory for Condensed Matter Physics, and Institute of Physics, Chinese Academy of Sciences, Beijing 100190, China}
\author{Z. Fang}
\affiliation{Beijing National Laboratory for Condensed Matter Physics, and Institute of Physics, Chinese Academy of Sciences, Beijing 100190, China}
\author{Z. Wang}
\affiliation{Department of Physics, Boston College, Chestnut Hill, MA 02467, USA}
\author{H. Ding}\email{dingh@bc.edu}
\affiliation{Department of Physics, Boston College, Chestnut Hill, MA 02467, USA}
\affiliation{Beijing National Laboratory for Condensed Matter Physics, and Institute of Physics, Chinese Academy of Sciences, Beijing 100190, China}

\date{\today}% It is always \today, today,
ÊÊÊÊÊÊÊÊÊÊÊÊ% Êbut any date may be explicitly specified

%\verb+\pacs{#1}+ command.

%\pacs{74.72.Jt, 74.25.Ha, 74.25.Ld}

%\keywords{Bi$_{2}$Sr$_2$CaCu$_2$O$_{8+x}$, ARPES, photon energy resonance, local state HTSC}
%Use showkeys class option if keyword display desired
\maketitle

{\bf
Electrons in a simple correlated system behave either as itinerant
charge carriers or as localized moments. However, there is growing 
evidence for the coexistence of itinerant electrons and local moments 
in transition metals with nearly degenerate $d$-orbitals. It demands one 
or more selective electron orbitals undergo the Mott transition while the 
others remain itinerant. Here we report the first observation of such an orbital 
selective Mott transition (OSMT) in Ca$_{1.8}$Sr$_{0.2}$RuO$_4$
by angle-resolved photoemission spectroscopy (ARPES). While we observed two
sets of dispersing bands and Fermi surface associated with the
doubly-degenerate $d_{yz}$ and $d_{zx}$ orbitals, the Fermi surface 
associated with the wider $d_{xy}$ band is missing, a consequence of selective 
Mott localization. Our theoretical calculations demonstrate that this novel OSMT 
is mainly driven by the combined effects of interorbital 
carrier transfer, superlattice potential, and orbital degeneracy, whereas the 
bandwidth difference plays a less important role.
}

Ca$_{2-x}$Sr$_x$RuO$_4$ is a fascinating 4$d$ multi-orbital system
that exhibits a rich and intricate phase diagram, ranging from a
chiral $p$-wave superconductor (Sr$_{2}$RuO$_4$) to a Mott insulator
(Ca$_{2}$RuO$_4$) \cite{Maeno,Phase_diag}. Similar to the high-$T_c$
cuprates, the metal-insulator transition under the influence of
electron correlations in the ruthenates is of fundamental importance
and currently under intensive debate. There is accumulating
experimental evidence for coexistence of local moments and metallic
transport, and heavy fermion behavior in the region of 0.2$\leq x
\leq$ 0.5 \cite{Phase_diag,Nakatsuji2003,Jin_crystal}, which is
remarkable because there are no $f$-electrons in this material. To
account for the coexistence of localized and itinerant electrons in
this region, a scenario of OSMT
has been proposed \cite{Anisimov,Koga} as following: Sr$_{2}$RuO$_4$
has three degenerated t$_{2g}$ orbitals ($d_{xy}$, $d_{yz}$, and
$d_{zx}$) occupied by four 4$d$ electrons. The isovalent Ca
substitute does not change the total carrier concentration (i.e. no
doping), but rather increases the effective electron correlation
strength ($U_{eff}$) relative to the reduced bandwidth which is
induced by structural change due to the smaller Ca$^{2+}$ ion
radius. Consequently, it is possible that an OSMT takes place in the
narrower bands, i.e., the one-deimensional (1D) $d_{yz}$ and
$d_{zx}$ orbitals, where electrons undergo a Mott transition and
become localized, while the electrons in the wider two-dimensional
(2D) $d_{xy}$ band remain itinerant. A similar partial localization
mechanism has been proposed for some heavy fermion materials, e.~g.,
UPd$_2$Al$_3$ \cite{SteglichNature}.

While the concept of ÊOSMT is of critical importance to
themulti-orbital Mott Hubbard systems and has been studied
extensively in theory
\cite{Anisimov,Koga,Liebsch_03,Liebsch_07,Fang}, it has not been
confirmed experimentally.
%While this type of OSMT is appealing, it has been challenged both theoretically \cite{Liebsch_03,Fang} and experimentally \cite{wang_PRL,neutron}.
Our previous ARPES \cite{wang_PRL} shows that, for samples with $x$ = 0.5, the two 1D ($\alpha$ and $\beta$) Fermi surface (FS) sheets are
clearly ``visible". The 2D ($\gamma$) FS, while being heavily smeared, survives at this substitution level. This is in sharp contrast to the OSMT prediction that the
1D FS sheets become Mott localized. To check if the OSMT occurs at a lower Sr concentration, we have conducted a series of ÊARPES experiments on high-quality single
crystals at $x$ = 0.2, grown by the floating zone technique \cite{Jin_crystal}. Precise determination of the low-energy electronic structure at this doping level is important since
Ca$_{1.8}$Sr$_{0.2}$RuO$_4$ is at the boundary between a magnetic metal and an antiferromagnetic insulator, and exhibits non-Fermi liquid behaviors in the resistivity \cite{Phase_diag}.
However, it is a rather difficult and lengthy experiment due to a much reduced ARPES spectral intensity in the vicinity of the Fermi energy ($E_F$) near the insulating phase.
Several techniques have been used to boost photoeletron signals,
including enhancement of the APRES matrix elements through fine tuning of photon energy and measurements at different Broullion zones (BZs). Ê

As shown in Fig.~1, strong spectral intensity and clear dispersion are observed in the valence bands of Ca$_{1.8}$Sr$_{0.2}$RuO$_4$, similar to the case of Sr$_{2}$RuO$_4$, indicating good
quality of sample and surface. However, the spectral intensity near $E_F$ experiences dramatic reductions as the Sr content $x$ approaches 0.2, as demonstrated in Fig.~1c,
reflecting the fact that the system is near an insulating phase. To see clearly the low-energy excitations, we zoom in to the binding energy range within 0.2 eV of $E_F$, as plotted in Fig.~2. One can
identify weak but discernible peaks dispersing towards to $E_F$, as shown in Figs.~2c and d. More specifically, we observe, as illustrated in Figs.~2a and b, two linearly dispersive bands along the high-symmetry line
$X$-$\Gamma$-$X$, or ($\pi$,$\pi$)-(0,0)-(-$\pi$,-$\pi$), crossing $E_F$ around 0.3 and -0.9 \AA$^{-1}$, respectively. These two Fermi crossing ($k_F$) points,
plotted in Fig.~2f as
points $\#$ 1 and 2, locate on the calculated $\alpha$ Fermi surface for Sr$_{2}$RuO$_4$ and its folded FS (labeled as the $\alpha\prime$ FS) due to the $\sqrt2\times\sqrt2$ reconstruction caused by
a rotation of the RuO$_6$ octahedra \cite{recon_Science}.
We note that the peak intensity in both energy distribution curves (EDCs) and momentum distribution curves (MDCs) diminishes as it approaches $E_F$ or $k_F$, possibly
due to a small energy gap or the quasiparticle decoherece effect observed in some transition metal oxides near the
metal-insulator boundary \cite{ghost_FS}.
The band dispersion along another high-symmetry line $M$-$\Gamma$-$M$ is displayed in Fig.~2E, and we observed four FS crossings (points $\#$3 - $\#$6)
whose locations are plotted in Fig.~2F. While the crossing points $\#$4 and 5 are on the calculated $\alpha\prime$ FS, and $\#$3 and 6 are close to the $\beta$ FS, there is no observation
of the dispersing $\gamma$ band and the corresponding FS crossing.

To verify these band and FS assignments, we have performed many measurements to cover a wide range of k-space. ÊWe locate and plot all observed FS crossing points in the first BZ
shown in Fig.~3a. It is clear that both the $\alpha$ and $\beta$ main FS sheets and the folded $\alpha\prime$ FS are present in Ca$_{1.8}$Sr$_{0.2}$RuO$_4$. However, no
evidence of the $\gamma$ FS is found. The disappearance of the $\gamma$ FS is very puzzling. According to Luttinger counting theorem, the total occupied FS area
should remain the same due to the isovalence nature of the Ca-Sr substitution. From the fitted $\alpha$ and $\beta$ FS sheets, as shown in Fig.~3a, we derive the electron occupations
$n_{\alpha} = 1.72$ and $n_{\beta} = 0.74$, implying that the $\gamma$ band has 1.52 valence electrons since $n_{total} = 4$. To illustrate this point, in Fig.~3a we plot the ``would be" $\gamma$
FS as a simple circle (black dished line), Êwhich satisfies the Luttinger counting of 1.52 electrons. Note that it would almost touch the $M$ ($\pi$, 0) point, indicating that its van Hove singularity is
very close to the Fermi energy, which may lead to instability at low temperature.

To further understand the fate of the $\gamma$ band, we plot in Fig.~3b an EDC of Ca$_{1.8}$Sr$_{0.2}$RuO$_4$ integrated over the neighboring region of $M$ (indicated by a rectangular
box around $M$ in Fig.~3a). In constrast to the EDC of Sr$_{2}$RuO$_4$ at $M$, which is also plotted in Fig.~3b, the EDC of Ca$_{1.8}$Sr$_{0.2}$RuO$_4$ shows a dramatic suppression of the $\gamma$
quasiparticles (QPs). In fact, the spectrum consists of a broad feature with a gap of $\gtrsim$ 100 meV, and a small ``foot" extending toward $E_F$. The origin of this small ``foot" is not entirely clear, although
it may possibly come from a residual $\gamma$ band from a minority phase, or certain impurity states. We regard the disappearance of the $\gamma$ QP with a large soft gap as an evidence for
possible localization of the $\gamma$ band. We notice that the electron occupancy of the $\gamma$ band is close to 1.5 (a half integer) from both the experimental derivation based on
Luttinger theorem as discussed above, and our theoretical calculation using local density approximation (LDA), as shown in Fig.~3c. It is remarkable that the LDA calculation shows good agreement with
the ARPES experimental observation. The basic reason for the increase of the $\gamma$ electron occupation is that the increased hybridization between the $t_{2g}$ and $e_{g}$ orbitals, due to
the increasing rotation and tilting of octahedra at higher Ca content, pushes down the $d_{xy}$ band. The same effect has been also observed in a similar 4$d$-electron system (Sr$_2$RhO$_4$) \cite{Kim}.

A natural question is why the FS and the coherent excitations from
the $\gamma$ band are absent at 1.5 electron occupancy. Remarkably,
as the system undergoes the $\sqrt2\times\sqrt2$ reconstruction in
the bulk \cite{recon_Science}, the $\gamma$ band folds into two
subbands by the superlattice potential, accompanied by the doubling
of the unit cell. The folded $\gamma$ bands in the reduced
Brillouin zone host a total of 3 electrons. The lower subband is
completely filled while the upper one is precisely at half-filling. It
is thus possible for the $\gamma$-complex to undergo the Mott
transition and become localized, contributing a spin-1/2 local
moment. Since there are two Ru atoms per supercell, the localized
magnetic moment is 0.5 $\mu_{B}$ per Ru atom. This is indeed
consistent with the field dependent magnetization measurement in
Ca$_{1.8}$Sr$_{0.2}$RuO$_4$ \cite{Nakatsuji2003}. The sharp increase
of the magnetization to 0.5 $\mu_{B}$ as the applied magnetic field
reaches about 5T can be attributed naturally to the polarization of
the local moment, whereas the subsequent gradual growth of the
magnetization with further increasing field arises from Pauli
paramagnetism of the itinerant $\alpha$ and $\beta$ band electrons.

%In the following, we provide the theoretical support for the above
%picture within the framework of OSMT.
Theoretically, most of the model
studies have focused on the two-band Hubbard model, where the OSMT
is mainly controlled by the difference in the bandwidths
\cite{Koga,Liebsch_03}. The real situation in $Ca_{2-x}Sr_{x}RuO_4$
system is more complex. Due to the $\sqrt2\times\sqrt2$ super
structure, there is a total of six bands occupied by eight electrons
in the doubled unit cell.
%The LDA calculation indicates that the
%widths of the $\alpha$, $\beta$ and $\gamma$ bands are similar and,
%as we will discuss below, the crystal field splitting plays a very
%important role in this system.
We have carried out first-principle
calculations and found that the lower three bonding bands are fully
occupied by six electrons. The remaining two electrons occupy the
upper three anti-bonding bands. If there were no crystal field
splitting, these two electrons would be almost evenly distributed
among the upper three bands, corresponding to occupations
$(2/3,2/3,2/3)$. However, the localized
orbital must be filled by an odd integer number of electrons in an OSMT, which
can be realized in $Ca_{2-x}Sr_{x}RuO_4$ only when the two electrons
redistribute among the upper three bands to reach occupations
$(1/2,1/2,1)$ due to the crystal field splitting. This is
consistent with our ARPES measurements near $x=0.2$. Indeed, our LDA
calculation shows that, with the reduction of the $Sr$ concentration
$x$, the crystal field pulls down the $\gamma$ band and transfers
charge from the $\alpha$, $\beta$ bands to the $\gamma$ band as shown in
Fig.~3c. When the electron distribution reaches $(1/2,1/2,1)$, the
three-band complex splits into two groups: two nearly degenerate
$\alpha$ and $\beta$ bands, and a separated $\gamma$ band with a
lowered center of gravity. Therefore, we have one two-band system
and one single band system, both with one electron per unit cell.
The reason for the OSMT to take place in the $\gamma$ band is that
the critical interaction $U_c$ for the Mott transition in a single
band Hubbard model is about 30\% smaller than that of a two-band
model with one electron per unit cell and identical bandwidth, a
result obtained by variational Gutzwiller and dynamical meanfield
theory \cite{multi-orb}. The Hund's rule coupling further increases
the critical $U_c$ for the two-band system. Therefore,
%if the correlation between the two systems is weak, it is reasonable to expect that
in a large area of the parameter space, the single
$\gamma$-band system lies in the Mott phase contributing the local
moment, while the two-band system remains in the metallic phase
contributing itinerant electrons.
%Thus, the charge transfer from the
%$\alpha$, $\beta$ band to the $\gamma$ band reduces the inter-band
%correlations, pushes the occupation toward $(1/2,1/2,1)$, and
%induces the OSMT in $Ca_{2-x}Sr_{x}RuO_4$ near $x=0.2$.

To further illustrate this point, we apply the slave boson mean
field theory to a simple three-band Hubbard model with the bandwidth
ratios of 1:1:1 and 1:1:1.5 and two electrons per unit cell. The
technical details have been explained in reference
\cite{slave-boson}. In Fig.~4, we plot the quasiparticle coherence
weight ($Z$) and the orbital correlations as a function of the
charge transfer $\delta$. Fig.~4a clearly shows that the coherence
weight of the $\gamma$ band decreases continuously while that of the
degenerate $\alpha$ band remains almost a constant as the OSMT is
approached at charge transfer $\delta=1/3$, which corresponds to the
charge distribution $(1/2,1/2,1)$. Concomitantly, as can be seen
from Fig.~4b, the inter-band correlations ($\chi$) between the
$\gamma$ and the $\alpha$, $\beta$ bands are dramatically reduced.
To verify that the bandwidth difference does not play an important
role in the OSMT, we show in Figs.~4c and 4d that the same
conclusion is reached for the case where the $\gamma$ band is 1.5
times as wide as that of the $\alpha$ and $\beta$ bands.
%These results provide strong theoretical support for the multi
%orbital correlation physics discussed above, and the mechanism of
%the OSMT in $Ca_{2-x}Sr_{x}RuO_4$ observed by our ARPES
%measurements.

In conclusion, we have successfully measured the low energy
excitations in multi orbital ruthenate $Ca_{2-x}Sr_{x}RuO_4$ near
$x=0.2$ by ARPES and unraveled a novel mechanism for the OSMT. The
low-energy band dispersions and the associated FS are observed for
the $d_{yz}$ and $d_{zx}$ orbitals. In contrast, the $d_{xy}$
orbital shows a loss of coherent low energy quasiparticle
excitations due to Mott localization. We discovered that the
$\sqrt{2}\times\sqrt {2}$ structure reconstruction plays a crucial
role in establishing the half-filling condition of the anti-bonding
$\gamma$ band. We provided microscopic theoretical support for this
novel OSMT and demonstrated the importance of the crystal field
splitting induced interorbital charge transfer and the orbital
degeneracy for promoting an intriguing electronic phase with
coexisting local moment and itinerant electrons. Our findings
highlight the emergent and fundamentally important phenomena
governed by the Mott physics in multi-orbital correlated electron
systems, and call for more systematic studies of transition metal
based materials.

\section*{METHODS}

All of our experiments have been performed at high-flux
synchrotron undulator beam lines (e.g., Wadsworth,U1-NIM,PGM at the Synchrotron Radiation Center, Wisconsin), using a high-efficiency Scienta SES-2002 electron analyzer. The energy and
momentum resolutions are 10 - 30 meV and 0.02 \AA$^{-1}$, respectively. Samples were
cleaved {\it in situ} and measured at 40 K in a vacuum better than 1 x 10$^{-10}$ torr. The samples have been found to be very stable and without degradation for
the typical measurement period of 48 hours.

%\bibliography{biblio_en2}

\section*{Acknowledgments}
This work was supported by grants of US NSF DMR-0353108,
DMR-0704545, US DOE DEFG02-99ER45747, and China NSF. This work is
based upon research conducted at the Synchrotron Radiation Center
supported by NSF DMR-0537588. Oak Ridge National laboratory is
managed by UT-Battelle, LLC, for DOE under contract
DE-AC05-00OR22725.

\section*{Author information}
Correspondence and requests for materials should be addressed to H.D.

\pagebreak

%***********************************Figure 1**************************************
\begin{figure}[htbp]
\begin{center}
\includegraphics[width=14cm]{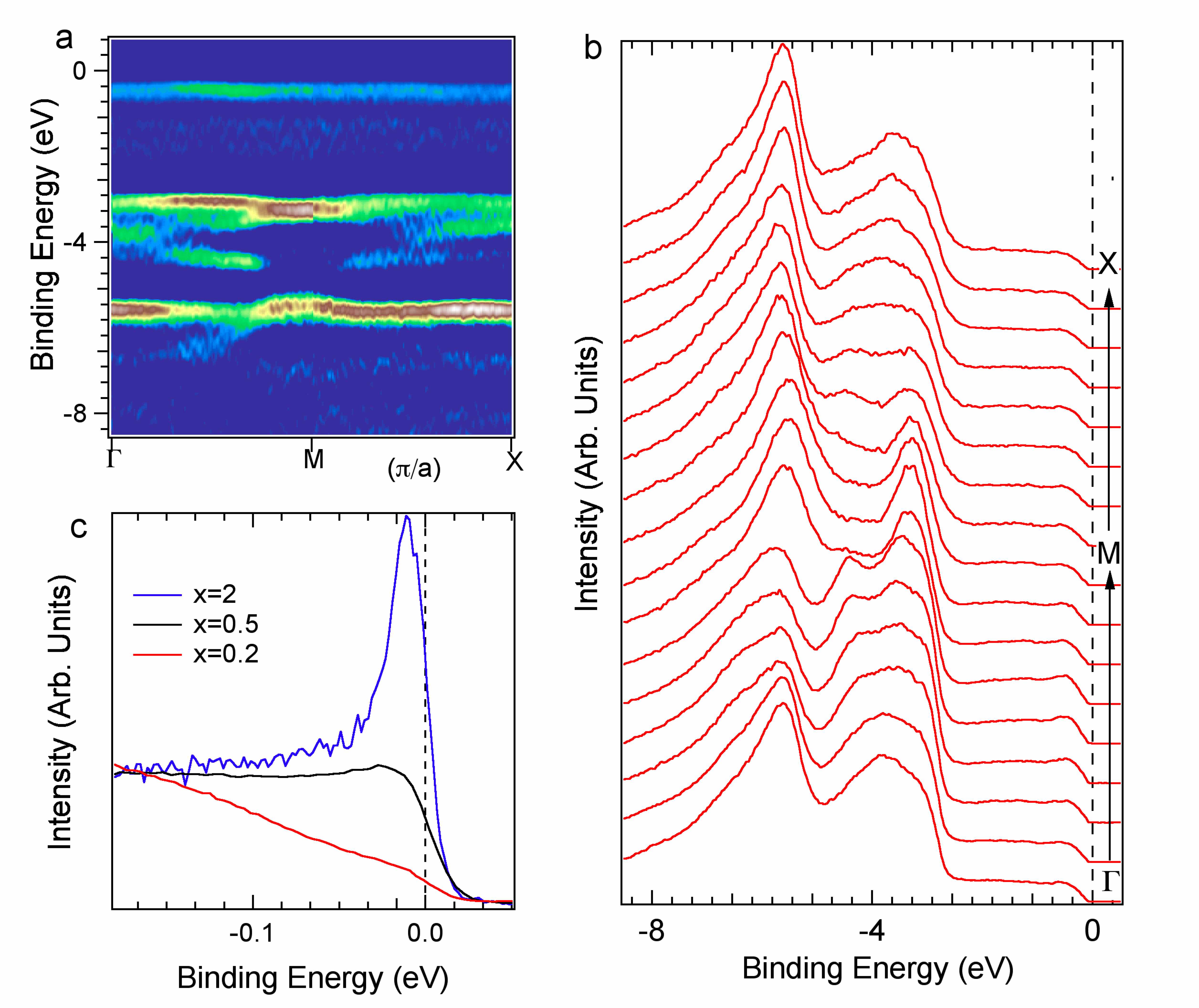}
\caption{\label{Figure1} {\bf Valence bands of Ca$_{1.8}$Sr$_{0.2}$RuO$_4$.} (a) Plot of second derivative of ARPES intensity for the valence band of Ca$_{1.8}$Sr$_{0.2}$RuO$_4$ taken along $\Gamma$-$M$-$X$ Ê(h$\nu$ = 75 eV, $T$ = 40 K). (b) The corresponding EDCs along $\Gamma$-$M$-$X$. (c) Comparison of the EDCs for $x$ = Ê0.2, 0.5, 2, taken at the $\beta$ Êband Êcrossing point along $\Gamma$-$M$ (h$\nu$ = 32 eV, $T$ = 40 K)}.
\end{center}
\end{figure}

\pagebreak

%***********************************Figure 2**************************************
\begin{figure}[htbp]
\begin{center}
\includegraphics[width=14cm]{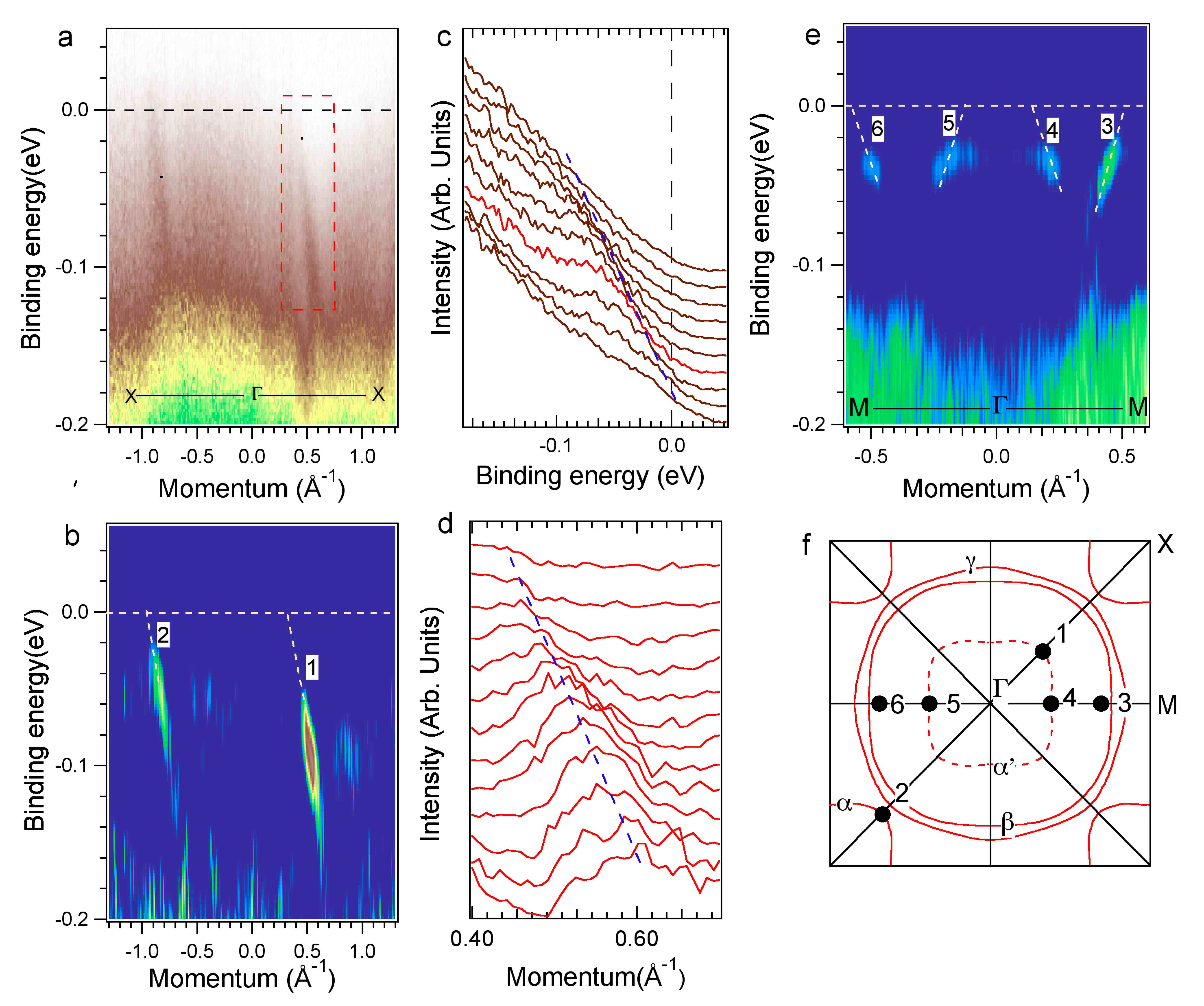}
\caption{\label{Figure 2} {\bf Band dispersion along high-symmetry directions in Ca$_{1.8}$Sr$_{0.2}$RuO$_4$.} Ê(a) ARPES intensity plot along $X$-$\Gamma$-$X$. (b) The corresponding second derivative plot of (a). The white dashed lines are linear extrapolations for the band dispersion. (c) EDCs, and (d) MDCs, within the red dashed box in (A), showing low-energy band dispersion, as indicated by blue dashed lines. (e) Second derivative plot of ARPES intensity along $M$-$\Gamma$-$M$. (f) Fermi crossings (black dots) indicated in (b) and (e), and Fermi surface sheets (red lines) calculated by LDA for Sr$_{2}$RuO$_4$ \cite{LDA} in the first BZ.}
\end{center}
\end{figure}

\pagebreak

%***********************************Figure 3**************************************
\begin{figure}[htbp]
\begin{center}
\includegraphics[width=11cm]{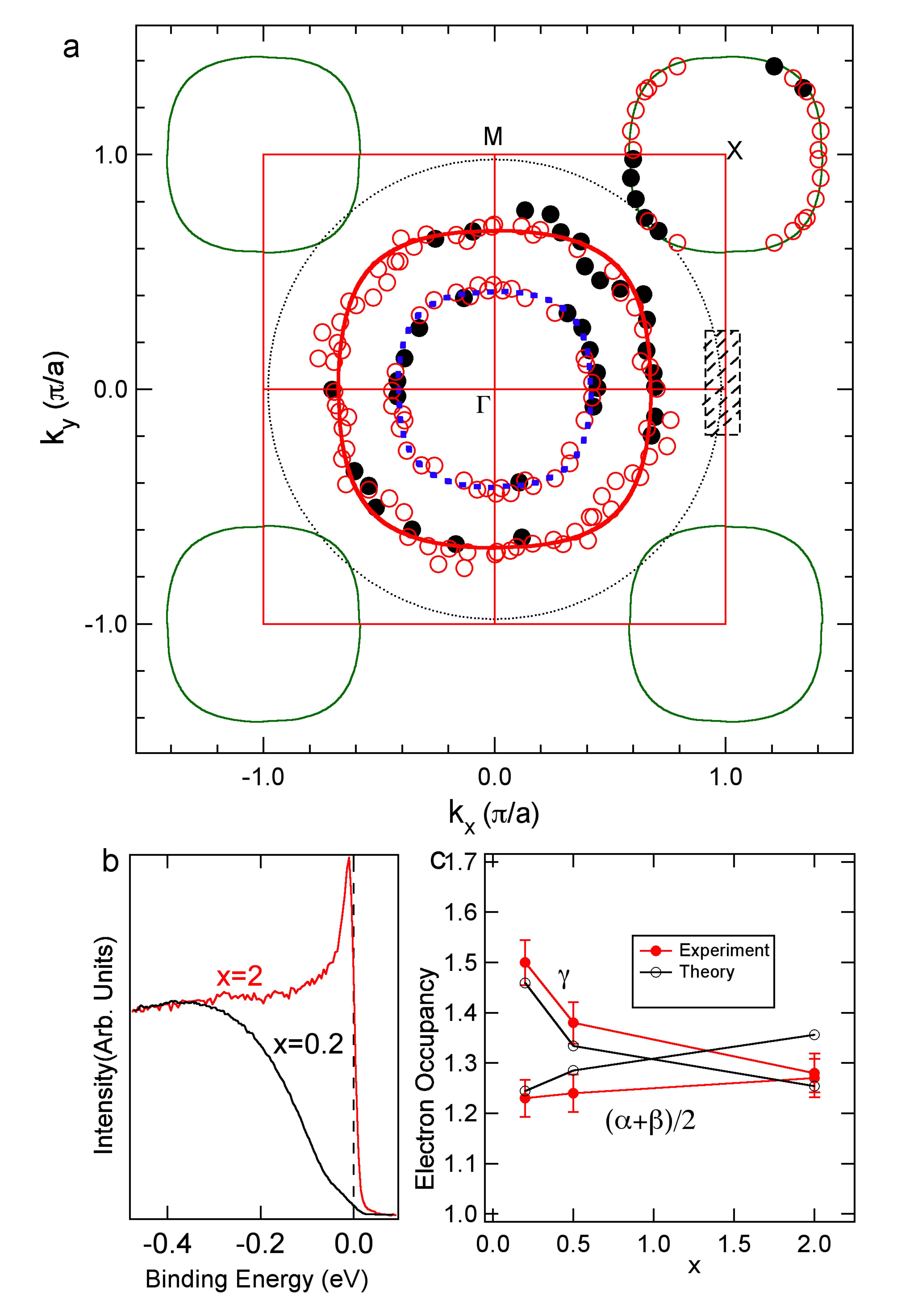}
\caption{\label{Figure 3} {\bf Fermi surface and electron occupancy of different orbitals.} (a) Measured Fermi surface sheets of $\alpha$ (green contours centered at $X$), $\beta$ (red contour centered at $\Gamma$), and the folded $\alpha$ (blue dashed contour centered at $\Gamma$), along with the Fermi crossing points determined by ARPES measurements (black solid dots) and symmetrized points according to the 4-fold crystal symmetry (red open dots). The black dotted contour centered at $\Gamma$ is the derived $\gamma$ Fermi surface according to Luttinger counting theorem. (b) Comparison of the ($\pi, 0$) EDCs between Ca$_{1.8}$Sr$_{0.2}$RuO$_4$ and Sr$_{2}$RuO$_4$, integrated over the $k$-region indicated by the shaded rectangle in (a). (c) The Sr content dependence of electron occupancy of $\gamma$ and ($\alpha$+$\beta$)/2, obtained by ARPES measurements (red dots and lines), and LDA calculations (black dots and lines).
}
\end{center}
\end{figure}

\pagebreak

%***********************************Figure 4**************************************
\begin{figure}[htbp]
\begin{center}
\includegraphics[width=12cm]{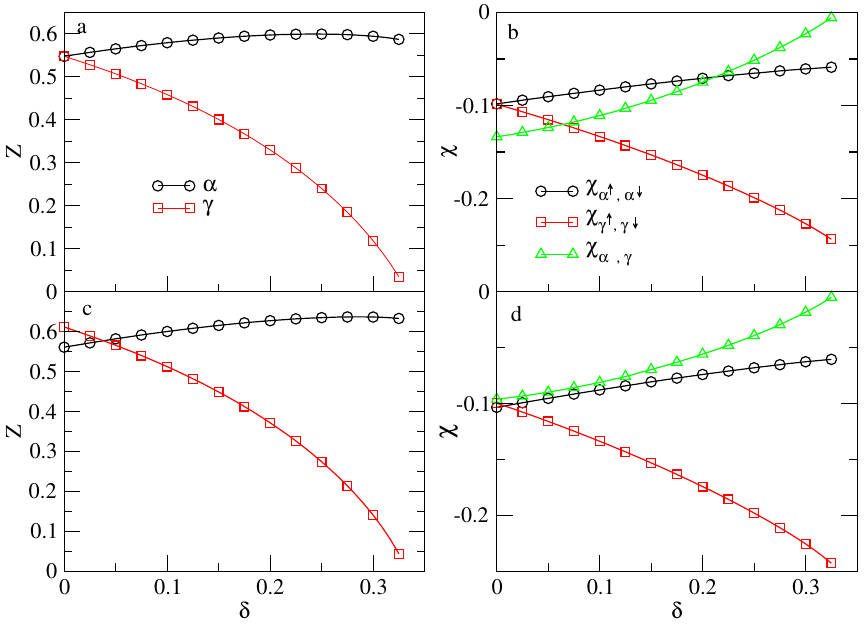}
\caption{\label{Figure 4} {\bf Calculations that show OSMT in multi-orbital systems}
(a) The quasiparticle weight for the
3-band Hubbard model with equal bandwidth W as a function of charge
transfer $\delta$ defined by ($n_\alpha$, $n_\beta$, $n_\gamma$) =
($\frac{2}{3}$-$\frac{\delta}{2}$, $\frac{2}{3}$-$\frac{\delta}{2}$,
$\frac{2}{3}$+$\delta$). U/W=4.0, J/W=1.0. (b) The intra-band and
inter-band correlation functions, where
$\chi_{\alpha{\uparrow},\alpha{\downarrow}}=<n_{\alpha{\uparrow}}n_{\alpha{\downarrow}}>
-<n_{\alpha{\uparrow}}><n_{\alpha{\downarrow}}>$,
$\chi_{\gamma{\uparrow},\gamma{\downarrow}}=<n_{\gamma{\uparrow}}n_{\gamma{\downarrow}}>
-<n_{\gamma{\uparrow}}><n_{\gamma{\downarrow}}>$ and
$\chi_{\alpha,\gamma}=<(n_{\alpha{\uparrow}}+n_{\alpha{\downarrow}})(n_{\gamma{\uparrow}}+n_{\gamma{\downarrow}})>
-<(n_{\alpha{\uparrow}}+n_{\alpha{\downarrow}})><(n_{\gamma{\uparrow}}+n_{\gamma{\downarrow}})>$.
(c) Same as in panel A, except for the three bandwidths of
$W,W,1.5W$. (d) Same as in panel b, except for the three bandwidths
of $W,W,1.5W$. Note that the y-axis is negative in (b) and (d). With
increasing $\delta$, the intra $\gamma$-band correlation becomes
large and negative, while the interband correlation approaches zero.
}
\end{center}
\end{figure}

\pagebreak

\end{document}